\documentclass[aps,prb,showpacs,twocolumn,superscriptaddress]{revtex4-1}
\usepackage{graphicx,amssymb,amsmath,bm,color,flushend}

\newcommand{\jav}[1]{\textcolor{black}{#1}}

\begin{document}
\definecolor{darkgreen}{rgb}{0,0.5,0}
\definecolor{matlabmagenta}{rgb}{1,0,1}

\title{Defect production due to time-dependent coupling to environment in the Lindblad equation}

\author{Bal\'azs Gul\'acsi}
\email{gulacsi@phy.bme.hu}
\affiliation{Department of Theoretical Physics and MTA-BME Lend\"ulet Topology and Correlation Research Group,Budapest University of Technology and Economics, 1521 Budapest, Hungary}
\author{Bal\'azs D\'ora}
\affiliation{Department of Theoretical Physics and MTA-BME Lend\"ulet Topology and Correlation Research Group,Budapest University of Technology and Economics, 1521 Budapest, Hungary}
\date{\today}

\begin{abstract}
Recently defect production was investigated during non-unitary dynamics due to non-Hermitian Hamiltonian.
By ramping up the non-Hermitian coupling linearly in time through an exceptional point, defects are produced in much the same way as approaching a Hermitian critical point.
A generalized Kibble--Zurek scaling accounted for the ensuing scaling of the defect density in terms of the speed of the drive and the corresponding critical exponents.
Here we extend this setting by adding the recycling term and considering the full Lindbladian time evolution of the problem with quantum jumps.
We find that by linearly ramping up the environmental coupling in time, and going beyond the steady-state solution of the Liouvillian, the defect density scales linearly with the speed of the drive for all cases.
This scaling is unaffected by the presence of exceptional points of the Liouvillian, which can show up in the transient states.
By using a variant of the adiabatic perturbation theory, the scaling of the defect density is determined exactly from a set of \emph{algebraic} equations.
\jav{Our study indicates the distinct sensitivity of the Lindbladian time evolution to exceptional points corresponding to steady states and transient states.}


\end{abstract}

\maketitle

\section{Introduction}

Over the recent years, enquiry for systems described by non-Hermitian Hamiltonians has greatly increased. Such systems are generically open in the sense that they are in touch with an environment.
Formal description of the dynamics of open quantum systems, like the Lindblad formalism\cite{lind},  consists of a Hermitian Hamiltonian  part, describing  the  coherent  evolution  of  the 
system and a non-Hermitian dissipator part, encompassing the loss of energy, information and coherence into the environment. These dissipators admit an interesting interpretation in  terms  
of  quantum  maps  and  measurement theory\cite{qmeas,Paris2012} and can be divided into two parts. The first one represents a coherent non-unitary dissipation of the  system, the second  
one describes quantum jumps, which are the effect of a continuous measurement performed by the environment on the system\cite{jump1}. 
These quantum jumps have been observed in countless 
experiments, including ionic\cite{ionexp1,ionexp2}, atomic\cite{atomexp1,atomexp2}, solid-state\cite{ssexp1,ssexp2} and
superconducting circuit setups\cite{scexp1,scexp2,scexp3}. For microscopic systems, such a quantum description is crucial, 
however macroscopic settings can be well described by semi-classical approaches in which the \jav{quantum noise due to the} jump terms are neglected\cite{elgana}. 
At this level, the process of quantum dissipation can be studied by the non-Hermitian effective Hamiltonian. Experimental progress on this subject 
are occurring in a wide range of different platforms such as photonics\cite{top1,Xiao2017,top2}, cold atomic systems\cite{CA1,CA2}, mechanical 
systems\cite{MS1,MS2} and electric circuits\cite{EC1,EC2,EC3}.

An important property of non-Hermitian dynamics is the existence of exceptional points (EPs). At an EP the complex spectrum of the Hamiltonian 
becomes gapless. These can be regarded as the non-Hermitian counterpart of conventional quantum critical points\cite{Heiss2012}. What is remarkable 
about EPs is that in these points the complex eigenvalues coalesce and the eigenstates no longer form a complete basis. This can lead to fascinating 
consequences in the dynamics whenever an EP is crossed.

Recently, universal defect production was predicted in a non-Hermitian many-body quantum system\cite{dora} during dynamics that cross an EP. Universal 
defect production in non-equilibrium many-body physics with unitary real-time evolution is understood by the Kibble--Zurek mechanism\cite{Kibble}. 
According to the adiabatic theorem, by sufficiently slowly changing a parameter in a Hamiltonian, the system can follow its ground state as long as
 the spectrum remains gapped. However, crossing a continuous phase transition during the dynamics will result in the development of excitations\cite{polk}. 
The scaling of these defects is universal as it is determined by the universality class of the underlying phase transition: 
\begin{gather}
n\sim\tau^{-\frac{d\nu}{z(\nu+1)}}.
\label{KZM}
\end{gather}
Here, $n$ stands for the number of defects, $\tau^{-1}$ denotes the rate at which the parameter is dynamically varied, $d$ the spatial dimension, $\nu$ and $z$ the correlation 
length and dynamical critical exponent, respectively. The rather peculiar observation is, that the defect density $n$ still obeys \jav{the same} scaling form in the non-Hermitian case\jav{, specifically crossing the EP during dynamics,} 
with the spatial dimension $d$ replaced by the effective dimension $d^*=d+z$. Furthermore, using single-photon interferometric networks the power-law dependence with exponents 
in accord  with  the  Kibble--Zurek  prediction were observed\cite{expkz}.

\jav{Of late, studies of critical quantum dynamics in the presence of dissipation due to interaction to the environment have gained considerable attention\cite{OKZM4,OKZM5,OKZM6}.
Numerical results for Kibble--Zurek protocols applied to a one-dimensional fermion wire undergoing a quantum phase transition, showed that open systems can develop universal dynamic scaling, which are controlled by the universality class of the quantum phase transition\cite{OKZM1}. 
Experimental tests of the Kibble--Zurek mechanism using D-wave quantum annealing devices have also indicated, that dynamical scaling predicted in isolated critical systems continue to hold in the presence of coupling to the environment\cite{OKZM2}. 
The open nature of quantum systems may also lead to a departure from the dynamic scaling behavior predicted for isolated systems. In particular, contact with an environment may break down the scaling laws and one may observe an anti-Kibble--Zurek behavior: slower ramps lead to less adiabatic dynamics, increasing thus non-adiabatic effects with the quench time\cite{OKZM3}.
We emphasize, the focus of these works were on the effects of the environment on the critical quantum dynamics, in contrast to the above mentioned non-Hermitian Kibble--Zurek dynamics, where criticality enters through the environment as an exceptional point.}


The motivation behind this work is to extend the study of the system examined in Ref.~\onlinecite{dora} from the effective Hamiltonian approach to the full Lindbladian 
dynamics containing quantum jumps. First, we identify the correct jump operators to construct the Liouvillian that describes the dynamics. Afterwards, we use coherence 
vector formalism to rewrite the Lindblad master equation into a matrix equation. By linearly ramping up the environmental coupling and going beyond the steady-state 
solution of the Liouvillian, the scaling of the defect production is calculated. We do this by expanding the density matrix into a series in terms of the slow speed of the drive. 
Remarkably, this series expansion does not break down even if the dynamics crosses the EPs of the Liouvillian. The convergence properties of the series reveals that
our exact solution is valid as long as the time scale introduced by  the linear ramp is longer than the other time scales in the problem.

\section{The model}
\label{sec2}

We start from the non-Hermitian Hamiltonians with the following form\cite{dora,sys}:
\begin{gather}
H_{eff}=\sum_pH_p,\quad H_p=p\sigma_x+\Delta\sigma_y+i\gamma\sigma_z,
\label{Heff}
\end{gather}
where $\sigma_x,\sigma_y,\sigma_z$ are the Pauli matrices and $p$ is the momentum. \jav{Its eigenvalues are $\pm\sqrt{p^2+\Delta^2-\gamma^2}$, which realize an EP 
for $p^2+\Delta^2=\gamma^2$.}
This Hamiltonian is interpreted as an effective Hamiltonian which originates from a Liouvillian without the quantum jump term.
 \jav{It can be regarded as the continuum limit of the famous Su--Schrieffer--Heeger (SSH) model
with $p$ the momentum,
$\Delta$ the alternating hopping,
 supplemented by  alternating imaginary, non-Hermitian on-site energies~\cite{longhi} $\gamma$. Our effective Hamiltonian also describes the
quantum Ising chain in an imaginary transverse field.}
According to the Lindblad formalism, the state of the system is described by a density operator and its time evolution is governed by the Liouvillian:
\begin{gather}
\frac{d\rho}{dt}=\mathcal L\rho.
\end{gather}
The Liouvillian is written as\cite{lind}:
\begin{gather}
\mathcal L\rho=-i[H,\rho]+\Gamma\rho\Gamma^\dagger-\frac{1}{2}\{\rho,\Gamma^\dagger\Gamma\},
\label{lindi}
\end{gather}
where the $\Gamma$ operators are called quantum jump operators and they describe the interaction between the system with Hamiltonian $H$ and the environment. It is straightforward to show that Eq.~\eqref{lindi} can be rewritten to
\begin{gather}
\mathcal L\rho=-i(H_{eff}^\dagger\rho-\rho H_{eff})+\Gamma\rho\Gamma^\dagger,
\label{lindi2}
\end{gather}
with $H_{eff}=H+\frac{i}{2}\Gamma^\dagger\Gamma$. Comparing this to Eq.~\eqref{Heff} we immediately recognize that apart from a constant term, that has no effect on the dynamics, they are the same if the jump operator is chosen as
\begin{gather}
\Gamma=\sqrt{\gamma}(\sigma_x-i\sigma_y)=2\sqrt\gamma\sigma_-.
\end{gather}
\jav{Eq. \eqref{lindi} is interpreted as the continuum limit of the dissipative SSH-model with alternating balanced  gain and loss processes.}

In Eq.~\eqref{lindi2} the last term $\Gamma\rho\Gamma^\dagger$ is called the recycling or quantum jump term. 
\jav{The label quantum jump originates from a quantum trajectory approach, i.e. the description of the open system by a stochastic wavefunction\cite{jump2}. 
In this formalism, the stochastic wavefunction smoothly evolves according to the non-Hermitian effective Hamiltonian, Eq. \eqref{Heff} until 
it abruptly changes under the action of the environment, which is controlled by the recycling term.
The Lindblad master equation describes the average over infinite quantum trajectories. In some of these experiments, quantum jumps took place. 
The semi-classical approach simply counts only those experiments where no quantum jump happened. If such description is sufficient, the dynamics 
can be studied by the properties of the non-Hermitian effective Hamiltonian.}
Henceforward, our goal is to describe the full Lindbladian dynamics of the system with Hamiltonian
\begin{gather}
H=\sum_pH_p,\quad H_p=p\sigma_x+\Delta\sigma_y,
\label{Ham}
\end{gather}
which is connected to the environment via the jump operator $\Gamma=2\sqrt{\gamma}\sigma_-$. 
\jav{The 
Hamiltonian in Eq.~\eqref{Ham} is the 1D gapped Dirac-equation and describes two level systems labeled with momentum $p$ with an energy 
gap $\Delta$.} 
The eigenenergies and eigenvectors of the system are  
\begin{gather}
E_{\pm}=\pm\sqrt{p^2+\Delta^2},\quad|\psi_\pm\rangle=\frac{1}{\sqrt{2(p^2+\Delta^2)}}\begin{pmatrix}
E_\pm(p)\\p+i\Delta
\end{pmatrix}.
\label{epm}
\end{gather}


Our main focus is on the effect of linear ramping $\gamma\sim t/\tau$, starting from $t=0$.
The initial state at $t=0$ is considered to be the ground-state of $H_p$. We will see however, that the long time dynamics of the system is independent from the initial condition.
In Ref.~\onlinecite{dora}, two main settings were investigated using only $H_{eff}$, without the quantum jump term.
In the first case, $\Delta$ was finite and $\gamma=\Delta t/\tau$ until $t=\tau$,  while in the second case, $\Delta=0$ and $\gamma=\gamma_0t/\tau$ until $t=\tau$.
In the first case, an EP is reached at the end of the time evolution, while
the second case features the continuous movement of an EP in momentum space.
Since the time dependent $\gamma(t)$ couples to $\sigma_z$ in Eq. \eqref{Heff}, the expectation value of $\sigma_z$ qualifies as defect density, similarly to
other problems~\cite{polk,dziarmagareview}.
After the time evolution,
the expectation value of $\sigma_z$ was evaluated, and the defect density, measured from the adiabatic ($\tau\rightarrow\infty$) 
limit was found to scale as $\tau^{-2/3}$ and $\tau^{-1}$, respectively.
\jav{These scalings follow from the non-Hermitian Kibble-Zurek mechanism\cite{dora} according to Eq.~\eqref{KZM}, 
with the spatial dimension replaced by the effective dimension $d^*=d+z$ and $d=1$. The critical exponents of Eq.~\eqref{Heff} with $\Delta\neq0$ are 
$z=1,\ \nu=1/2$, which give the $\tau^{-2/3}$ scaling of the defect density. Similarly, the critical exponents of Eq.~\eqref{Heff} with 
$\Delta=0$ are $z=1/2,\ \nu=1$, giving $\tau^{-1}$ scaling.}
Here we investigate the fate of the corresponding quantum jump term and full Lindbladian time evolution on defect production.

\begin{figure}[h]
\includegraphics[scale=.6]{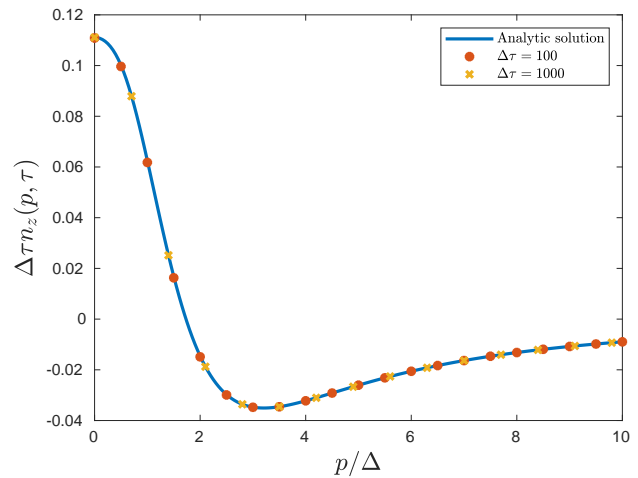}
\caption{The defect production of the gapped system $(\Delta\neq0)$ scaled with $\tau$ and $\gamma_0=\Delta$. The solid line depicts the analytical solution from 
Eq.~\eqref{exact}, the red dots and yellow crosses correspond to the numerical solution of Eq.~\eqref{sys} with $\Delta\tau=100$ and $\Delta\tau=1000$, respectively.}
\label{num}
\end{figure}

To this end, the numerical solution of the Lindblad equation reveals that on top of  the steady-state expectation  value of the Pauli
matrices ($\sigma_i^{(ss)}$), the defect production scales as
\begin{gather}
\langle\sigma_i(p,\tau)\rangle-\sigma_i^{(ss)}=n_i(p,\tau)\sim\frac{1}{\tau},
\end{gather}
where $\langle\sigma_i(p,\tau)\rangle$ denotes the expectation value of the Pauli matrices, evaluated from the time dependent density matrix at the end of the protocol.
The very same scaling holds for the above two non-Hermitian protocols, supplemented with quantum jumps, depicted in Figs. \ref{num} and \ref{nogap}.
In the following we show that this scaling can be understood by the series expansion of the density matrix in terms of $1/\tau$.
\jav{Within the same formalism, we also investigate the Lindblad equation \emph{without} the recycling term in the Appendix and find, that the non-Hermitian Kibble-Zurek scaling
of Ref. \onlinecite{dora} is indeed reproduced. The difference of the scaling in the presence or absence of the recycling term is traced back to the distinct asymptotic properties of the 
series expansion of the density matrix  in terms of $1/\tau$.}

\subsection{The coherence-vector formalism}  
Equipped with the correct form of the jump operator the Lindblad master equation can be studied:
\begin{gather}
\frac{d\rho}{dt}=-i[H_p,\rho]+4\gamma(t)\left(\sigma_-\rho\sigma_+-\frac{1}{2}\{\rho,\sigma_+\sigma_-\}\right).
\label{sys}
\end{gather}
As was mentioned, the environmental interaction is linearly ramped up: $\gamma(t)=\gamma_0t/\tau$, the precise value of $\gamma_0$ will be chosen to match the protocols of Ref.~\onlinecite{dora}. 

It is imperative to use coherence-vector formalism\cite{cvf} and represent the density operator as a vector. In our case this is just the typical Bloch vector representation of the density operator as we are dealing with two level systems. We write the density operator as a $2\times2$ matrix:
\begin{gather}
\rho=\frac{1}{2}(\mathbb I+\mathbf v\boldsymbol\sigma),
\end{gather}
where $\mathbb I$ is the unit matrix and $\mathbf v$ is the Bloch vector. The components of the Bloch vector is actually the average of the 
quantities represented by the Pauli matrices: $v_i=$Tr$(\rho\sigma_i)$. Since in the space of complex $2\times2$ matrices, the Pauli matrices along with the unit matrix form a basis, the density operator can be represented as a four component vector: $
|\rho\rangle=\frac{1}{2}(1\ v_1\ v_2\ v_3)^T$. This formalism allows us to rewrite the Lindblad master equation into the matrix equation $|\dot\rho\rangle=L(t)|\rho\rangle$, where the Liouvillian is represented as the supermatrix
\begin{gather}
L(t)=\begin{pmatrix}
0&&0&&0&&0\\
0&&-2\gamma(t)&&0&&2\Delta\\
0&&0&&-2\gamma(t)&&-2p\\
-4\gamma(t)&&-2\Delta&&2p&&-4\gamma(t)
\end{pmatrix}.
\label{matrix}
\end{gather}
From the matrix equation we can numerically study the dynamics for different ramping protocols, see Fig.~\ref{num}.

\subsection{Properties of the Liouvillian supermatrix}
Let us analyze the properties of the Liouvillian first. The matrix in Eq.~\eqref{matrix} is obviously non-Hermitian. This results in the fact that its right and left eigenvectors are not the Hermitian conjugate of each other, they have to be calculated independently\cite{adiab}. 
The instantaneous eigenvalues are however the same for left ($\langle\mathcal E_\beta(t)|$)  and right ($|\mathcal D_\alpha(t)\rangle$)
eigenvectors:
\begin{gather}
L(t)|\mathcal D_\alpha(t)\rangle=\lambda_\alpha(t)|\mathcal D_\alpha(t)\rangle,\nonumber\\
\langle\mathcal E_\beta(t)|L(t)=\langle\mathcal E_\beta(t)|\lambda_\beta(t).
\end{gather}
With a little algebra, it can be shown that for different eigenvalues the left and right eigenvectors are biorthogonal: $\langle\mathcal E_\beta(t)|\mathcal D_\alpha(t)\rangle=0$, if $\beta\neq\alpha$. Due to the non-Hermiticity the eigenvalues are not necessarily real:
\begin{gather}
\lambda_0(t)=0,\lambda_1(t)=-2\gamma(t),\nonumber\\\lambda_{2,3}(t)=-3\gamma(t)\pm i\sqrt{4(p^2+\Delta^2)-\gamma(t)^2}.
\label{eigv}
\end{gather}
\jav{Compared to the spectrum of Eq. \eqref{Heff} in the absence of the recycling term, these eigenvalues do not give rise to EP for the same set of parameters as in Ref. \onlinecite{dora}. 
Consequently, the recycling term washes away the criticality present in $H_{eff}$ in Eq. \eqref{Heff}. 
However, Eqs. \eqref{eigv} possess an EP for
the transient states, $\lambda_{2,3}$ at $4(p^2+\Delta^2)=\gamma^2$. As we demonstrate in Sec. \ref{secep}, even when the dynamics passes through 
this EP, it does not affect the ensuing scaling of the defect density.}

\begin{figure}[h]
\includegraphics[width=8cm,height=6cm]{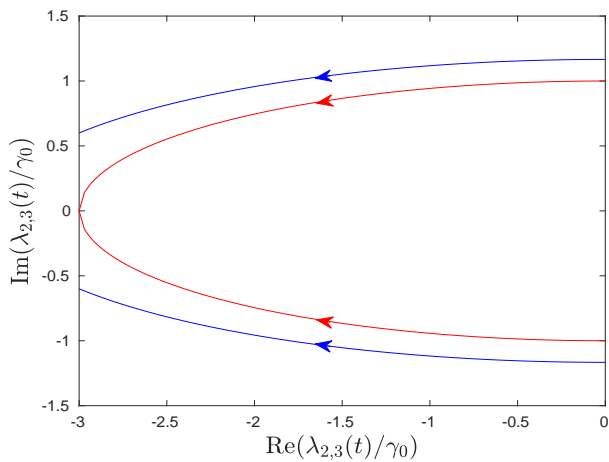}
\caption{The $\lambda_{2,3}$ eigenvalues of the Liouvillian for $\Delta\neq0$ and $\gamma(t)=\gamma_0t/\tau$, with $\gamma_0=2\Delta$ as time evolves from $0\to\tau$ depicted by the arrows on the curves. The red curves denote the eigenvalues for $p=0$, as we can see at the end of the evolution they coalesce. The blue curves denote the eigenvalues whenever $p\neq0$. If $\gamma_0=\Delta$ the eigenvalues will look like the blue curves for every $p$, signing the absence of EPs.}
\label{sajat}
\end{figure}

We can observe, that the first row of $L(t)$ are zeroes. As a consequence the first component of the right eigenvectors corresponding to nonzero eigenvalues is  zero. In our representation this means that $|D_{\alpha\neq0}\rangle$ corresponds to traceless $2\times2$ matrices, which in turn means that they cannot be associated with a density operator. However, after spectral decomposition they can be related to physical states\cite{spect}. The zero eigenvalue and the corresponding right eigenvector is connected to the steady-state solution of the dynamics. The Liouvillian has a unique steady state due to its single zero eigenvalue. The system always evolves into this steady state regardless of the initial state.
Similarly, the left eigenvectors of nonzero eigenvalues have zero as the first component. Furthermore, the left eigenvector corresponding to the zero eigenvalue    must have zeroes in its second, third and fourth component:
\begin{gather}
\langle\mathcal E_0(t)|=\begin{pmatrix}
1&&0&&0&&0
\end{pmatrix}.
\end{gather}

\jav{As was mentioned, if the environmental coupling reaches $2\Delta$ the Liouvillian will exhibit an EP corresponding to the transient states at $p=0$, see Fig.~\ref{sajat}.} Keeping the consistency with the PT symmetric ramp of Ref.~\onlinecite{dora} and varying the coupling until $\gamma=\Delta$ means that the dynamics never reaches this EP. However, we will discuss the situation when the dynamics reaches the EP. Furthermore, in the case when the gap $\Delta$ is zero $(\Delta=0)$, there is an EP whenever $p=\gamma/2$. Thus, we will study the gapless case separately.

\section{Defect production}
The real part of the eigenvalues of the Liouvillian apart from $\lambda_0$ are negative, hence $|\mathcal D_0\rangle$ contains information about the steady-state expectation values of the quantities represented by the Pauli matrices.  Since we are interested in the defect production during the ramping of the environmental coupling, we should look for a solution of the density matrix as a linear combination of the right instantaneous eigenvectors:
\begin{gather}
|\rho(t)\rangle=\sum_{\alpha=0}^3r_\alpha(t)|\mathcal D_\alpha(t)\rangle.
\label{ro1}
\end{gather}
As it was mentioned $|D_{\alpha\neq0}\rangle$ corresponds to traceless matrices, to keep consistency in the formalism $|D_0\rangle$ has to be normalized so that Tr$\rho=1$. 
This normalization reveals the aforementioned steady-state expectation values:
\begin{gather}
|\mathcal D_0(t)\rangle=\frac{1}{2}\begin{pmatrix}
1\\ \frac{-2\Delta\gamma}{p^2+\Delta^2+2\gamma^2}\\ \frac{2p\gamma}{p^2+\Delta^2+2\gamma^2} \\\frac{-2\gamma^2}{p^2+\Delta^2+2\gamma^2}
\end{pmatrix}=
\frac{1}{2}\begin{pmatrix}
1\\ \sigma_x^{(ss)}\\ \sigma_y^{(ss)} \\\sigma_z^{(ss)}
\end{pmatrix},
\end{gather} 
and the time dependence enters through $\gamma(t)$, what we dropped here to ease notations.
Since this takes care of the normalization of the density matrix, it is not necessary to normalize the rest of the eigenvectors:
\begin{widetext}
\begin{gather}
|\mathcal D_1(t)\rangle=\begin{pmatrix}
0\\ \frac{p}{\Delta}\\ 1 \\0
\end{pmatrix},\quad 
\langle\mathcal E_1(t)|=\begin{pmatrix}
0&&\frac{p}{\Delta}&&1&&0
\end{pmatrix},\quad
|\mathcal D_{2,3}(t)\rangle=\begin{pmatrix}
0\\ \frac{-\Delta(\gamma\pm i\sqrt{4(p^2+\Delta^2)-\gamma^2})}{2(p^2+\Delta^2)}\\ \frac{p(\gamma\pm i\sqrt{4(p^2+\Delta^2)-\gamma^2})}{2(p^2+\Delta^2)} \\1
\end{pmatrix},\nonumber\\ \langle\mathcal E_{2,3}(t)|=\begin{pmatrix}
\frac{\gamma(3\gamma\pm i\sqrt{4(p^2+\Delta^2)-\gamma^2})}{(p^2+\Delta^2+2\gamma^2)}&&\frac{\Delta(\gamma\pm i\sqrt{4(p^2+\Delta^2)-\gamma^2})}{2(p^2+\Delta^2)}&&\frac{-p(\gamma\pm i\sqrt{4(p^2+\Delta^2)-\gamma^2})}{2(p^2+\Delta^2)}&&1
\end{pmatrix}.
\label{eigenvectors}
\end{gather}
\end{widetext}
If we make the time derivative of Eq.~\eqref{ro1}, use the Lindblad matrix equation and multiply the results with the left instantaneous eigenvector of $L(t)$, we arrive at a differential equation system that determines the coefficients $r_\alpha(t)$:
\begin{gather}
\dot r_\alpha=\lambda_\alpha(t) r_\alpha-\sum_{\beta=0}^3 \frac{\langle\mathcal E_\alpha(t)|\frac{d}{dt}|\mathcal D_\beta(t)\rangle}{\langle\mathcal E_\alpha(t)|\mathcal D_\alpha(t)\rangle}r_\beta.
\label{diffeq}
\end{gather} 
Here, we can see the appearance of the norms $\langle\mathcal E_\alpha|\mathcal D_\alpha\rangle$, furthering the unnecessity of normalization of these eigenvectors. 

The Lindblad master equation is trace preserving\cite{lind}, which carries over to Eq.~\eqref{diffeq} as $\dot r_0=0$. The eigenvectors of $\lambda_1$ are both time independent, hence $r_1$ is disconnected from the other coefficients: $\dot r_1=\lambda_1 r_1$. Thus, during time evolution $r_1$ is exponentially suppressed. The equations of $r_{2,3}$ are particularly important, since their linear combination characterize the defect production in the expectation values of the Pauli matrices:
\begin{gather}
n_x(p,\tau)=-\frac{\Delta\gamma(r_2+r_3)}{p^2+\Delta^2}-\frac{2i\Delta(r_2-r_3)}{p^2+\Delta^2}\sqrt{p^2+\Delta^2-\frac{\gamma^2}{4}},\nonumber\\
n_y(p,\tau)=\frac{p\gamma(r_2+r_3)}{p^2+\Delta^2}+\frac{2ip(r_2-r_3)}{p^2+\Delta^2}\sqrt{p^2+\Delta^2-\frac{\gamma^2}{4}},\nonumber\\
n_z(p,\tau)=2(r_2+r_3).
\label{def}
\end{gather}
Here the $\tau$ dependence enters through $r_{1,2}(\tau)$ and $\gamma(\tau)$ .

\subsection{Solving the equations of motion in the gapped case ($\Delta\neq0$)}
As we can see in Eq.~\eqref{def} the linear combinations of $r_2$ and $r_3$ determine the defect production during the dynamics. For this reason we are defining these linear combinations as new functions and rewrite the differential equation system in Eq.~\eqref{diffeq}. These combinations are:
\begin{gather}
P(t)=r_2(t)+r_3(t),\nonumber\\
M(t)=\frac{i}{\Delta}(r_2(t)-r_3(t))\sqrt{4(p^2+\Delta^2)-\gamma(t)^2}.
\label{comb}
\end{gather} 
The $1/\Delta$ is necessary for $P$ and $M$ to share dimensions.
Their differential equation system is constructed by combining Eq.~\eqref{eigv}, Eq.~\eqref{eigenvectors}, Eq.~\eqref{diffeq} and Eq.~\eqref{comb}:
\begin{gather}
\dot P=-3\gamma P+\Delta M+\frac{2\dot\gamma\gamma(p^2+\Delta^2)}{(p^2+\Delta^2+2\gamma^2)^2},\nonumber\\
\Delta\dot M=-3\gamma\Delta M-(\dot\gamma+4(p^2+\Delta^2)-\gamma^2)P\nonumber\\-\frac{2\dot\gamma(p^2+\Delta^2)(p^2+\Delta^2-\gamma^2)}{(p^2+\Delta^2+2\gamma^2)^2}.
\label{final}
\end{gather}
We are ramping the environmental coupling linearly with rate $\tau$, we write this generally as $\gamma(t)=\gamma_0t/\tau$. Therefore, the time scale is $\tau$ and we change the independent variable of the system from $t$ to $x=t/\tau$. The energy scale of the problem is the gap $\Delta$, hence we measure $\gamma_0$ in $\Delta$ units: $\gamma_0=\epsilon\Delta$, where $\epsilon\in\mathbb R$.  Furthermore, since $\hbar=1$ the momentum is also measured in $\Delta$ units: $y=p/\Delta$. The dimensionless equations of motion are
\begin{gather}
\frac{1}{\Delta\tau}P'=-3\epsilon xP+M+\frac{2\epsilon^2x(1+y^2)}{(\Delta\tau)(1+y^2+2\epsilon^2x^2)^2},\nonumber\\
\frac{1}{\Delta\tau}M'=-3\epsilon xM-(\frac{\epsilon}{\Delta\tau}+4(1+y^2)-\epsilon^2x^2)P\nonumber\\
-\frac{2\epsilon^2(1+y^2)(1+y^2-\epsilon^2x^2)}{(\Delta\tau)(1+y^2+2\epsilon^2x^2)^2}.
\end{gather}
The notation $P'$ means the derivative with respect to $x$. Because of the time derivative of $\gamma$ in Eq. \eqref{final}, the time scale $\tau$ appears by itself in these equations, which means
that we can look for a solution as a series expansion of this time scale, formally:
\begin{gather}
P(x,y,\tau)=\sum_{k=1}^{\infty}c_k(x,y)(\Delta\tau)^{-k},\nonumber\\
M(x,y,\tau)=\sum_{k=1}^{\infty}d_k(x,y)(\Delta\tau)^{-k}.
\label{expansion}
\end{gather}
\jav{In the Appendix, we demonstrate that this expansion can break down in the absence of the recycling term.}
Finding the expansion coefficients $c_k$ and $d_k$ is achieved by plugging Eq.~\eqref{expansion} into their differential equations and balancing the multiplier of $(\Delta\tau)^{-k}$ for every $k$. This leads to a recursive formula for the coefficients, which are the following in the gapped case:
\begin{gather}
c_1(x,y)=\frac{\epsilon^2(1+y^2)((\epsilon^2+3\epsilon)x^2-y^2-1)}{2(1+y^2+2\epsilon^2x^2)^3},\nonumber\\
d_1(x,y)=\nonumber\\\frac{\epsilon^2x(1+y^2)((9\epsilon^2+3\epsilon^3-8\epsilon^4)x^2-(3+4\epsilon^2)y^2-(3+4\epsilon^2))}{2(1+y^2+2\epsilon^2x^2)^3},\nonumber\\
c_{k}=\frac{-3\epsilon xc'_{k-1}-d'_{k-1}-\epsilon c_{k-1}}{4(1+y^2)+8\epsilon^2x^2},\nonumber\\
d_{k}=\frac{(4(1+y^2)-\epsilon^2x^2)c'_{k-1}-3\epsilon xd'_{k-1}-3\epsilon^2x c_{k-1}}{4(1+y^2)+8\epsilon^2x^2},
\end{gather}
where $k\geq2$.
The expansion of the density matrix is complete. The expectation values of the Pauli matrices, the Bloch vector, can be calculated at $t=\tau$ by substituting $x=1$ into to formulas of $P(x,y,\tau)$ and $M(x,y,\tau)$. For example the expectation value of the defect production of $\sigma_z$ in the gapped system, with $\epsilon=1$ is
\begin{gather}
\langle n_z(p,\Delta,\tau)\rangle=\frac{\Delta(p^2+\Delta^2)(3\Delta^2-p^2)}{\tau(p^2+3\Delta^2)^3}+\mathcal O(\tau^{-2}).
\label{exact}
\end{gather}
In Fig.~\ref{num} the analytic curve, the depiction of Eq.~\eqref{exact}, fits well with the numerical data. Integrating with respect to the momentum $p$ reveals the defect density to be
\begin{gather}
\langle n_z(\tau)\rangle=\int_{-\infty}^\infty \frac{\text d p}{2\pi}\ \langle n_z(p,\Delta,\tau)\rangle=-\frac{1}{12\sqrt 3 \tau}+\mathcal O(\tau^{-2}).
\end{gather}
The analysis of the expansion in Eq.~\eqref{expansion} reveals the series converges if the magnitude of the time scale is large compared to other time scales, i.e. 
$\Delta\tau\gg10$, see Fig.~\ref{ck}. We remark, that the above results are independent from the initial conditions as the series solution of the equations are asymptotic.

\subsection{Solving the equations of motion in the gapless case $(\Delta=0)$}
If $\Delta=0$ the energy scale comes from the environmental coupling: $\gamma(t=\tau)=\gamma_0$ and the momentum is measured in units of $\gamma_0$. Also, keep in mind that the definition of $M$ in Eq.~\eqref{comb} changes to keep $P$ and $M$ the same dimension: $1/\Delta\to1/\gamma_0$. The dimensionless equations in the gapless case become
\begin{gather}
\frac{1}{\gamma_0\tau}P'=-3xP+M+\frac{2xy^2}{(\gamma_0\tau)(y^2+x^2)^2},\nonumber\\
\frac{1}{\gamma_0\tau}M'=-3xM-(\frac{1}{\gamma_0\tau}+4y^2-x^2)P\nonumber\\
-\frac{2y^2(y^2-x^2)}{(\gamma_0\tau)(y^2+2x^2)^2}.
\end{gather}
The appearance of the time scale by itself again motivates us to look for a solution in a form of an expansion:
\begin{gather}
P(x,y,\tau)=\sum_{k=1}^{\infty}c_k(x,y)(\gamma_0\tau)^{-k},\nonumber\\
M(x,y,\tau)=\sum_{k=1}^{\infty}d_k(x,y)(\gamma_0\tau)^{-k},
\label{expansion1}
\end{gather}
Balancing the equations of motion for every $k$ reveals the recursive formula for the coefficients in the gapless case:
\begin{gather}
c_1(x,y)=\frac{y^2(4x^2-y^2)}{2(y^2+2x^2)^3},\nonumber\\
d_1(x,y)=\frac{xy^2(4x^2-7y^2)}{2(y^2+2x^2)^3},\nonumber\\
c_{k}=\frac{-3xc'_{k-1}-d'_{k-1}-c_{k-1}}{4y^2+8\epsilon^2x^2},\nonumber\\
d_{k}=\frac{(4y^2-x^2)c'_{k-1}-3xd'_{k-1}-3x c_{k-1}}{4y^2+8x^2},
\label{nogapexp}
\end{gather}
where again $k\geq2$. The expectation value of the defect production of $\sigma_z$ in the gapless system is
\begin{gather}
\langle n_z(p,\tau)\rangle=\frac{\gamma_0 p^2(4\gamma_0^2-p^2)}{\tau(p^2+2\gamma_0^2)^3}+\mathcal O(\tau^{-2}),
\end{gather}
with defect density
\begin{gather}
\langle n_z(\tau)\rangle=-\frac{1}{16\sqrt 2 \tau}+\mathcal O(\tau^{-2}).
\end{gather}
\begin{figure}[h]
\includegraphics[scale=.6]{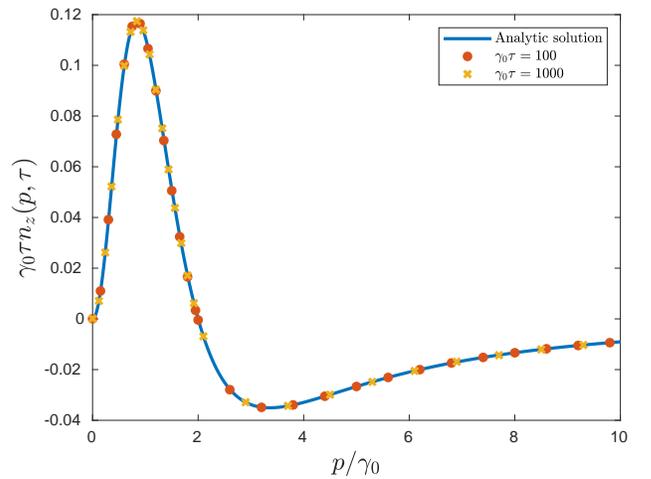}
\caption{Comparing $c_1$ from Eq.~\eqref{nogapexp} to the numerical solution of Eq.~\eqref{sys}. The solid line depicts the analytical solution, the red dots and yellow crosses correspond to the numerical solution with $\gamma_0\tau=100$ and $\gamma_0\tau=1000$, respectively. }
\label{nogap}
\end{figure}

\begin{figure}[h]
\includegraphics[scale=.6]{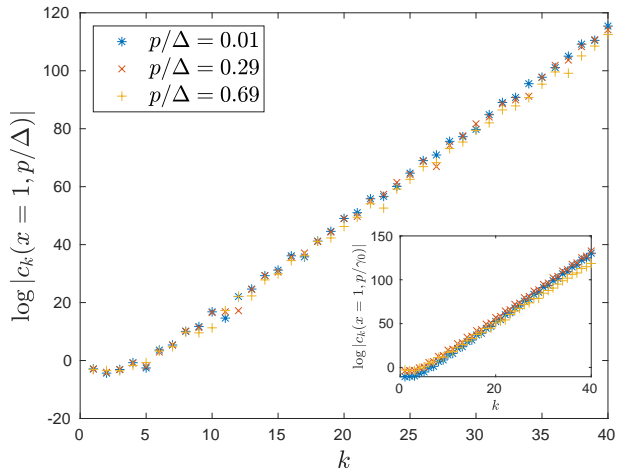}
\caption{For different momenta, the $c_k$ coefficients increase exponentially with $k$. The figure shows the coefficients in the gapped system, the inset shows the coefficients in the gapless system.  From the slope of the curves, we can conclude that in order for the series from Eq.~\eqref{expansion} and Eq.~\eqref{expansion1} to converge the rate has to have 
magnitude $\Delta\tau\gg10$ for the gapped case and $\gamma_0\tau\gg10$ for the gapless case.}
\label{ck}
\end{figure}

Comparing these results to the numerical data shows, that the expansion is indeed correct, see Fig.~\ref{nogap}. The analysis of the expansion reveals the order of the convergence radius 
to be $\gamma_0\tau\gg10$, see Fig.~\ref{ck}. Again, these results are independent from the initial conditions of the system.

\subsection{The role of the EP}
\label{secep}
\begin{figure}[h]
\includegraphics[scale=0.6]{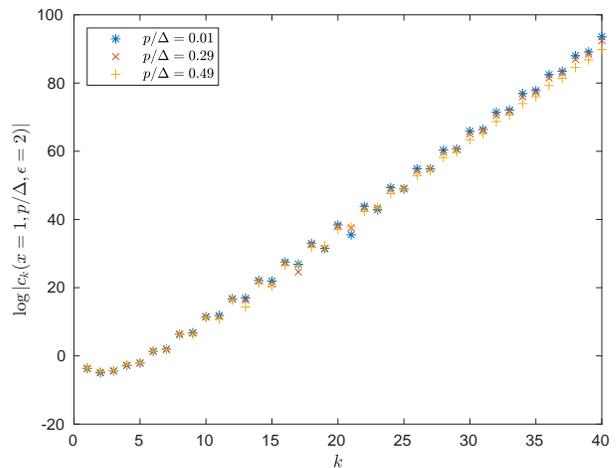}
\caption{\jav{The figure shows the coefficients in the gapped system for dynamics that reaches the EP: $\epsilon=2$. For different momenta, the $c_k$ coefficients increase exponentially with $k$ the same way as in for dynamics that does not reach the EP. We can conclude, that the presence of an EP in the transient states has no effect on the dynamics.} }
\label{ck2}
\end{figure}

As was discussed, in the presence of a gap, the dynamics crosses an EP of the Liouvillian 
whenever $\epsilon\geq2$. Remarkably, the expansion of the density matrix does not break 
down at the EP, see Fig.~\ref{ck2}. Thus, the EP has no effect on the dynamics. Furthermore, if a gap is not present the Liouvillian exhibits an EP if $p=\gamma/2$. Again, the expansion of density matrix does not break down at these exceptional points, thus in this system the EPs have no effect on the dynamics regardless of the presence of a gap.

\section{Discussion and Conclusion}
We have investigated the fate of the defect production in a one dimensional system coupled to the environment, which is described by the Lindblad equation. 
Linearly ramping up the coupling  by driving rate $1/\tau$ will result in defect production scaling as $\sim\tau^{-1}$ for all cases. 
We showed that this scaling is explained by a series expansion of the density matrix in terms of the driving rate.

Comparing this result to Ref.~\onlinecite{dora}, we notice that describing the system with the non-Hermitian Hamiltonian from Eq.~\eqref{Heff} and 
ramping $\gamma$ in a PT symmetric way, which means $\Delta\neq0$ and $\gamma=\Delta t/\tau$, results in defect production with scale $\sim\tau^{-2/3}$. 
This scaling is traced back to a generalized Kibble--Zurek mechanism during the crossing of an EP of the non-Hermitian effective Hamiltonian. 
As we have shown here, the inclusion of the quantum jump term in the description of the dynamics results in a different defect production scaling.
\jav{As detailed in the appendix, one may replicate the effective Hamiltonian setting by studying Eq.~\eqref{lindi2} in the 
coherence vector formalism, with the absence of the recycling term. The resulting Liouvillian supermatrix exhibits an EP that 
corresponds to its steady state, in contrast to Eq.~\eqref{matrix} where the exceptional point appears for transient states. For quenches that do
 not reach the EP, the form of the density matrix adopts a series expansion in terms of the driving rate, implying linear defect production scaling. 
However, as soon as the dynamics reaches the EP the series expansion breaks down, and the $\tau^{-2/3}$ follows from the fact that the momentum also becomes scaled.
What is more, in the gapless case of the non-Hermitian effective Hamiltonian approach, when $\Delta=0$, the EP features a continuous movement in momentum 
space, hence it is always present. As a result, the density matrix cannot take the series expansion form and the momentum dependent density exhibits a 
scaled momentum always. Accidentally, after momentum integration the resulting defect scaling is linear and coincides with the scaling obtained from 
the full Lindbladian evolution containing the recycling term.}
 

In conclusion, the inclusion of the quantum jump term in the dynamics results in defect production which scales with the rate of the linear ramp as 
$\sim\tau^{-1}$ for both the gapped ($\Delta\neq0$) and the gapless ($\Delta=0$) Hamiltonians. 
\jav{ This scaling is unaffected by the EP of the Liouvillian, which corresponds to transient states and can be imagined as an analogue for an excited 
state quantum phase transition\cite{ESQPT}.} 
We arrived at these results by calculating the density
matrix exactly in a form of a series expansion in terms of the driving rate which is valid if the magnitude of the rate of the drive ($1/\tau$) is small compared to the other
energy scales that appear in the Lindblad equation.

\jav{Our results indicate, that the modified scaling follows from the change in the critical properties of the system. 
The inclusion of the quantum jump or recycling term moves the EP of the Liouvillian from the steady states to the transient states.  
Without reaching the EP in the dynamics, the density matrix can be expressed as a series expansion in terms of the driving rate, which 
results in linear defect production scaling. If the dynamics reaches an EP corresponding to the steady state, the non-Hermitian analogue 
of a quantum phase transition, the expansion breaks down and the defect production scaling is described by the non-Hermitian version of the Kibble--Zurek mechanism.  }

\flushend

\begin{acknowledgments}
his  research  is  supported  by  the  National  Research, Development   and   Innovation   Office–NKFIH   within   the Quantum Technology National Excellence Program (Project No.  2017-1.2.1-NKP-2017-00001),  K119442, K134437 and by the Romanian National Authority for Scientific Research and Innovation, UEFISCDI, under project no. PN-III-P4-ID-PCE-2020-0277.
\end{acknowledgments}
\flushend

\appendix*
\section{Dynamics without quantum jumps}

\jav{We replicate the effective Hamiltonian setting of Ref.~\onlinecite{dora}, 
by studying Eq.~\eqref{lindi2} in the coherence vector formalism, with the absence of the recycling term. The Liouvillian
\begin{gather}
\mathcal L\rho=-i(H_{eff}^\dagger\rho-\rho H_{eff}),
\label{lindi3}
\end{gather}
with $H_{eff}$ given in Eq.~\eqref{Heff}, does not conserve the trace of the density matrix. As a result, the coherence vector 
representation of the density matrix becomes $|\rho\rangle=\frac{1}{2}(v_0\ v_1\ v_2\ v_3)^T$, with $v_0$ containing the time dependence of the trace. 
The presence of this time dependent $v_0$ represents one of the major differences between the two cases with and without the recycling term.
Due to the time dependence of the trace, the expectation values of observables need to be normalized, for example $\langle\sigma_z\rangle=v_3/v_0$. 
The Lindblad matrix equation is $|\dot\rho\rangle=L(t)|\rho\rangle$, with the supermatrix of Eq.~\eqref{lindi3} being
\begin{gather}
L(t)=\begin{pmatrix}
0&&0&&0&&-2\gamma(t)\\
0&&0&&0&&2\Delta\\
0&&0&&0&&-2p\\
-2\gamma(t)&&-2\Delta&&2p&&0
\end{pmatrix}.
\label{matrix1}
\end{gather}
The eigenvalues of this supermatrix are
\begin{gather}
\lambda_0(t)=0,\ \lambda_1(t)=0,\nonumber\\\lambda_{2,3}(t)=\pm i\sqrt{p^2+\Delta^2-\gamma(t)^2}.
\label{eigv1}
\end{gather}
Here, we can observe that quenching the environmental interaction as $\gamma(t)=\Delta t/\tau$, results 
in an EP at the end of the dynamics with all eigenvalues becoming zero (for $p=0$). }

\jav{We proceed by identifying the left and right eigenvectors of Eq.~\eqref{matrix1} and look for a solution in the form of Eq.~\eqref{ro1}. The eigenvectors are
\begin{widetext}
\begin{gather}
|\mathcal D_0(t)\rangle=\begin{pmatrix}
1\\ -\frac{\gamma}{2\Delta}\\ \frac{\gamma}{2p} \\0
\end{pmatrix},\quad 
|\mathcal D_1(t)\rangle=\begin{pmatrix}
0\\\frac{p}{\Delta}\\1\\0
\end{pmatrix},\quad
|\mathcal D_{2,3}(t)\rangle=\begin{pmatrix}
\frac{\mp\gamma}{i\sqrt{p^2+\Delta^2-\gamma^2}}\\ \frac{\pm\Delta}{i\sqrt{p^2+\Delta^2-\gamma^2}}\\\frac{\mp p}{i\sqrt{p^2+\Delta^2-\gamma^2}}\\1
\end{pmatrix},\nonumber\\
\langle\mathcal E_0(t)|=\begin{pmatrix}
1&&\frac{\gamma}{\Delta}&&0&&0
\end{pmatrix},\quad
\langle\mathcal E_1(t)|=\begin{pmatrix}
0&&\frac{p}{\Delta}&&1&&0
\end{pmatrix},
\nonumber\\ \langle\mathcal E_{2,3}(t)|=\begin{pmatrix}
\frac{\mp\gamma}{i\sqrt{p^2+\Delta^2-\gamma^2}}&&\frac{\mp\Delta}{i\sqrt{p^2+\Delta^2-\gamma^2}}&&\frac{\pm p}{i\sqrt{p^2+\Delta^2-\gamma^2}}&&1
\end{pmatrix}.
\label{eigenvectors1}
\end{gather}
\end{widetext}
The differential equations for the coefficients $r_\alpha$ are obtained from Eq.~\eqref{diffeq}. 
The major difference from the full Lindbladian description is the role of $r_0$, which here obeys a 
non-trivial differential equation connected to $r_2$ and $r_3$. This is the direct consequence of the 
trace non-preserving nature of Eq.~\eqref{lindi3}. Similarly to Eq.~\eqref{comb}, we define the linear
 combinations of $r_2$ and $r_3$ as new functions: $P(t)=r_2(t)+r_3(t)$ and $M(t)=i(r_2(t)-r_3(t))$. 
Using the dimensionless variable $x=t/\tau$, and dimensionless parameter $y=p/\Delta$, the equations of motions are 
\begin{gather}
r_0'=\frac{x}{1+y^2-x^2}r_0-\frac{1+y^2}{(1+y^2-x^2)^\frac{3}{2}}M,\nonumber\\
P'=2(\Delta\tau)\sqrt{1+y^2-x^2}M,\nonumber\\
M'=\frac{-1}{\sqrt{1+y^2-x^2}}r_0-2(\Delta\tau)\sqrt{1+y^2-x^2}P.
\label{diffeq2}
\end{gather}
In principle, we can look for a solution in a series expansion form, similarly to Eq.~\eqref{expansion}:
\begin{gather}
P(x,y,\tau)=\sum_{k=1}^{\infty}c_k(x,y)(\Delta\tau)^{-k},\nonumber\\
M(x,y,\tau)=\sum_{k=1}^{\infty}d_k(x,y)(\Delta\tau)^{-k},\nonumber\\
r_0(x,y,\tau)=\sum_{k=0}^{\infty}e_k(x,y)(\Delta\tau)^{-k}.
\label{expansion2}
\end{gather}
}

\begin{figure}[h]
\includegraphics[scale=.6]{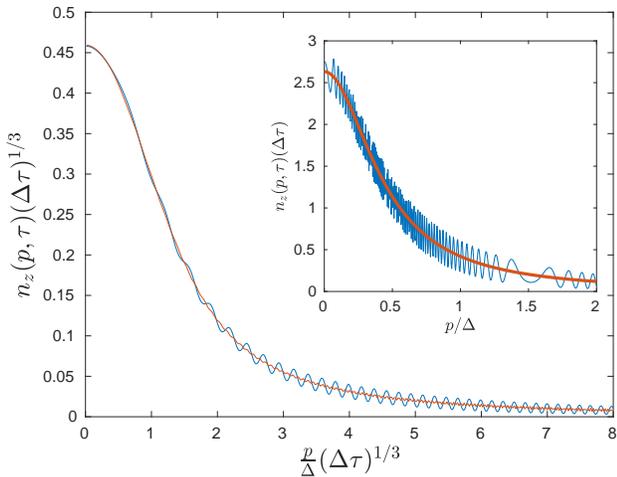}
\caption{\jav{The momentum dependent defect density obtained from the numerical solution of Eq.~\eqref{diffeq2}. The main plot shows the 
dynamics reaching the EP, with scaling in agreement with Eq.~\eqref{scaleapp}, for $\Delta\tau=10^2$ and $\Delta\tau=10^3$. The inset 
shows dynamics that does not reach the EP and a scaling in agreement with the series expansion solution: solid blue, the numerical sol
ution of Eq.~\eqref{matrix1} for $\Delta\tau=10^3$ and $x=0.95$; solid red, the analytical result based on the series expansion coefficients.}}
\label{app1}
\end{figure}
\jav{The expansion coefficients are obtained after balancing the multipliers of $(\Delta\tau)^{-k}$ for every $k$, which leads to a recursive formula for $k\geq1$:
\begin{gather}
c_k(x,y)=-\frac{e_{k-1}(x,y)}{2(1+y^2-x^2)}-\frac{d'_{k-1}(x,y)}{2\sqrt{1+y^2-x^2}},\nonumber\\
d_k(x,y)=\frac{c'_{k-1}(x,y)}{2\sqrt{1+y^2-x^2}},\nonumber\\
e_k(x,y)=e_0(x,y)-\frac{1+y^2}{2\sqrt{1+y^2-x^2}}\int\text d x\ \frac{c'_{k-1}(x,y)}{\sqrt{1+y^2-x^2}},
\end{gather}
where $c_0=d_0=0$ and $e_0(x,y)=const/\sqrt{1+y^2-x^2}$. The recursive formula reveals that for dynamics 
reaching the EP, i.e. $x=1$, each coefficient is divergent as a function of $y$ at the EP, as is evident from e.g. $e_0(1,0)$. 
The expansions in Eq.~\eqref{expansion2} assume the scaling to be independent from the momentum. However, in Ref.~\onlinecite{dora}, 
it was explained that the momentum is also scaled in the expectation value of the momentum dependent defect density. The numerical 
solution of Eq.~\eqref{diffeq2}, as well as Eq.~\eqref{matrix1} agrees with Ref.~\onlinecite{dora} and replicates the scaling properties of the momentum dependent defect density as
\begin{gather}
\langle n_z(p,\Delta,\tau)\rangle=(\Delta\tau)^{-1/3}f\left(\frac{p}{\Delta}(\Delta\tau)^{1/3}\right),
\label{scaleapp}
\end{gather}
with $f(x)$ being the universal scaling function depicted in Fig.~\ref{app1}.
After momentum integration, the scaling predicted by the non-Hermitian Kibble--Zurek mechanism follows as $n_z\sim\tau^{-2/3}$. }

\jav{These results indicate, the assumption of a scaling independent momentum breaks down for dynamics reaching the EP, and as a 
consequence the series expansion in Eq.~\eqref{expansion2} becomes invalid. However, for dynamics that does not reach the EP, i.e. $x<1$, 
the expansion coefficients display convergent behaviour. As a result, the numerical solution of Eq.~\eqref{diffeq2}, as well as Eq.~\eqref{matrix1} 
agrees with the scaling independent momentum dependence predicted by the series expansions in Eq.~\eqref{expansion2}, depicted in the inset of Fig.~\ref{app1}.}

\bibliographystyle{apsrev}
\bibliography{refgraph}

\end{document}